\documentclass[aps, physrev, reprint, twocolumn, superscriptaddress]{revtex4-2}

\usepackage{graphicx}
\usepackage{amsmath}
\usepackage{amssymb}
\usepackage{bm}
\usepackage{url}
\usepackage{hyperref}

\usepackage[capitalise]{cleveref}
\crefname{figure}{Fig.}{Figs.}
\Crefname{figure}{Fig.}{Figs.}

\usepackage{siunitx}
\newcommand{\qtyadj}[2]{\qty[number-unit-product={\text{-}}]{#1}{#2}}
\DeclareSIUnit{\molar}{M}
\usepackage[utf8]{inputenc}
\usepackage[T1]{fontenc}
\usepackage{lmodern}
\hypersetup{
	colorlinks=false,
	pdfborder={0 0 0},
    pdftitle={Microscopy of Bioelectric Potentials using Electrochromism},
    pdfauthor={Burhan Ahmed, Erica Liu, Lothar Maisenbacher, Pengwei Sun, Dana Griffith,
               Kenneth Nakasone, Ashwin Singh, Yuecheng Zhou, Bianxiao Cui, Holger Müller},
	pdflang={en-US}
}

\newcommand{\DRR}{\Delta R/R}
\newcommand{\DRRpp}{(\Delta R/R)_\mathrm{pp}}

\newcommand\newR[1]{{\color{blue}#1}}

\begin{document}

\title{\textbf{Microscopy of Bioelectric Potentials using Electrochromism}}

\author{Burhan Ahmed}
\affiliation{Department of Physics, University of California, Berkeley, California, USA}

\author{Erica Liu}
\affiliation{Department of Chemistry, Stanford University, Stanford, California, USA}
\affiliation{Wu Tsai Neurosciences Institute, Stanford University, Stanford, California, USA}

\author{Lothar Maisenbacher}
\affiliation{Department of Physics, University of California, Berkeley, California, USA}

\author{Pengwei Sun}
\affiliation{Department of Radiology, Stanford University, Stanford, California, USA}
\affiliation{Wu Tsai Neurosciences Institute, Stanford University, Stanford, California, USA}

\author{Dana Griffith}
\affiliation{Department of Physics, University of California, Berkeley, California, USA}

\author{Kenneth Nakasone}
\affiliation{Department of Physics, University of California, Berkeley, California, USA}

\author{Ashwin Singh}
\affiliation{Department of Physics, University of California, Berkeley, California, USA}

\author{Yuecheng Zhou}
\affiliation{Department of Materials Science and Engineering, The Grainger College of Engineering, University of Illinois Urbana-Champaign, Urbana, Illinois, USA}
\affiliation{Department of Bioengineering, The Grainger College of Engineering, University of Illinois Urbana-Champaign, Urbana, Illinois, USA}

\author{Bianxiao Cui}
\affiliation{Department of Chemistry, Stanford University, Stanford, California, USA}
\affiliation{Wu Tsai Neurosciences Institute, Stanford University, Stanford, California, USA}

\author{Holger Müller}
\email{Contact author: hm@berkeley.edu}
\affiliation{Department of Physics, University of California, Berkeley, California, USA}
\affiliation{Molecular Biophysics and Integrated Bioimaging, Lawrence Berkeley National Laboratory, Berkeley, California, USA}

\date{January 30, 2026}

\begin{abstract}

Studying the electrical signals generated by living cells is key to understanding numerous biological phenomena. Electrochromic optical recording (ECORE) uses the electrochromism exhibited by certain materials to noninvasively measure these signals in real time. In this work, we report on the development of ECORE based on a high-NA microscope objective. We demonstrate the recording of extracellular action potentials from cardiomyocytes with single-cell resolution and a high sensitivity of \newR{\qty{6.3}{\micro\volt}}. Combining ECORE with microscopy simplifies the optical setup, allows for the simultaneous imaging of specimens, and makes ECORE accessible to a broader community of researchers, allowing for a better understanding of the biological processes that are integral to life.
\end{abstract}

\maketitle

\section{Introduction}

Detecting, recording, and monitoring bioelectric activities is essential for understanding a wide range of physiological functions such as memory formation \cite{miller_electrophysiology, memory_formation}, muscle contraction \cite{skeletal_muscle_contraction}, and the beating of the heart \cite{cardiac_action_potentials}. In cardiomyocytes, for example, monitoring changes in the action potential can help identify the safety and efficacy of drugs \cite{cardiac, cardiac_drug_testing}. Methods to detect these potentials have been studied for many decades, with significant improvements over time to the signal-to-noise ratio and spatial resolution, among other aspects \cite{Lightelectrophysiology, voltage_imaging_neurons}. 

Electrode-based methods include the patch clamp \cite{patch_clamp_original, patchclamp}, which measures the large intracellular signal but is limited to short-term recording of very few recording sites due to its relatively large size and invasive nature. Measuring extracellular potentials by, e.g., multielectrode arrays \cite{mea_neurons, mea_method}, is less invasive and allows multichannel recording, but is limited to the fixed positions of the electrodes. 

Optical recording allows flexibility to choose multiple areas of interest in a specimen. Well-established methods use genetically-encoded fluorescent indicators that respond to bioelectric activities such as calcium concentration and voltage changes \cite{Lightelectrophysiology, voltageindicators}, but extended recording at a rapid rate is difficult due to photobleaching. Also, phototoxicity can compromise membrane integrity, while the introduction of the indicators can affect membrane capacitance \cite{voltageindicators}. Various label-free optical techniques have been proposed in response to these limitations \cite{Peterreview, afm, spr, nv_center, srs, interferometric}. 

One such label-free method is electrochromic optical recording (ECORE), which is based on the electrochromism \cite{electrochromism_review} in materials such as poly(3,4-ethylenedioxythiophene) polystyrene sulfonate (PEDOT:PSS, here referred to as PEDOT) \cite{PNAS, dualcolor} and other dioxythiophene-based polymers \cite{prodot}. In ECORE, cells are typically cultured on a PEDOT device (a glass slide coated with a thin film of PEDOT). Electrical potentials generated by the cells change the refractive index of the PEDOT film. Thus, by monitoring the power of a laser beam reflected off the device, these potentials can be measured. ECORE is noninvasive, allows recording from user-selected areas of interest, and is suitable for long-term recording: cells show no adverse reaction to PEDOT and can be measured over multiple recording sessions spanning many weeks \cite{PNAS, dualcolor, prodot}. ECORE recordings have a high sensitivity and have been shown to resolve electrical signals down to \qty{3.3}{\micro\volt}~\cite{prodot}, enabling the measurement of individual extracellular action potentials. Real-time recordings of action potentials associated with spontaneous activity from cardiomyocytes, neurons, and brain slices have already been demonstrated \cite{PNAS, dualcolor, prodot}.

In published ECORE setups, light is focused using a simple lens and directed towards the sample through a prism \cite{PNAS, dualcolor, prodot}. \newR{We use point-illumination of only selected sites of interest (as opposed to illuminating the whole sample) as it reduces the needed laser power and laser-induced heating. }This prism is coupled to the device slide using immersion oil; if the light is incident at the sample at supercritical angles, it is totally internally reflected and re-emerges from the other open face of the prism where it is sent to a detector. This geometry is challenging to align due to the oblique incidence angles and also requires separate imaging optics to observe the specimen. 

The incorporation of a microscope objective in ECORE is thus very desirable -- a high-numerical aperture (high-NA) objective can, in principle, simplify the alignment (since it can both direct the incident light to the sample at supercritical angles and collect the reflected light). The same objective can also be used for imaging, allowing simultaneous observation of a sample and the recording of its bioelectric potentials. However, the small reflectance changes (parts per thousand or lower) that need to be detected make it necessary to minimize technical noise, which has so far prevented the use of microscopes in ECORE. Noise from the interference of multiple light reflections in the objective (parasitic reflections), and the increased sensitivity to environmental vibrations due to the higher magnification and complexity of the objective, are of particular concern.

Here we report on microscope ECORE, which utilizes a high-NA total internal reflection objective to record bioelectric potentials, and use it to record spontaneous action potentials from cultured human induced pluripotent stem cell (hiPSC)-derived cardiomyocytes. Using a light source that is less susceptible to interference, along with vibration damping, we are able to record with high signal-to-noise ratios (SNRs) and achieve a detection limit of \newR{\qty{6.3}{\micro\volt}}. We use the same objective for imaging and show recordings of extracellular action potentials from different regions of one cardiomyocyte. By combining ECORE with microscopy, we have introduced a path to a broader implementation of ECORE for the label-free recording of bioelectric potentials.

\section{Experimental Setup}

\begin{figure*}
    \includegraphics{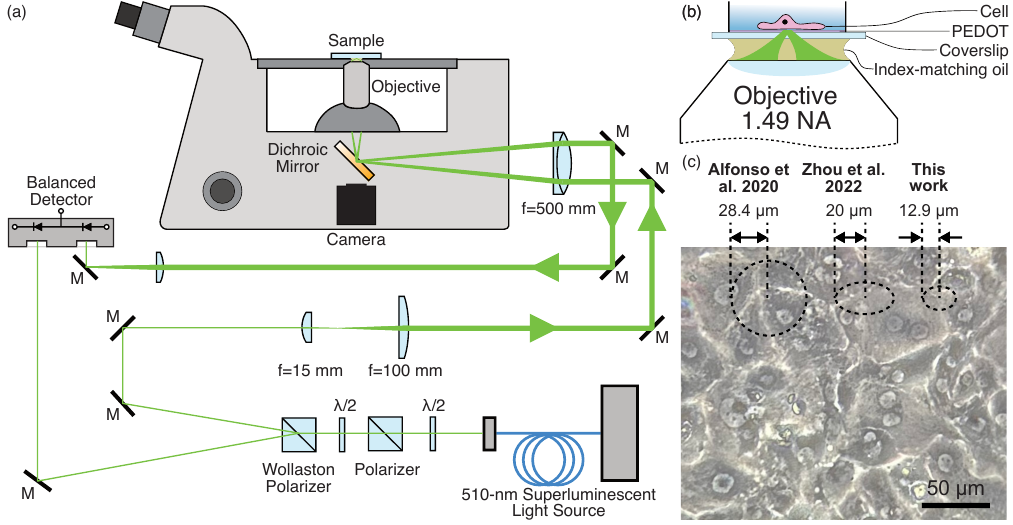}
    \caption{(a) Schematic drawing of microscope ECORE. M: mirror, $\mathrm{\lambda/2}$: half-wave plate. All components, except for the light source, are placed on an optical breadboard measuring $36 \times 24$ inches. (b) Detail showing the focused light at the sample. (c) A section of a microscope image (from a separate objective) showing a typical sample of cardiomyocytes. The dashed ellipses superposed on the image depict the $\mathrm{1/e^2}$ intensity radius spot sizes of various ECORE beams. From left to right: original ECORE \cite{PNAS}, dual-color ECORE \cite{dualcolor}, and microscope ECORE.}
    \label{fig:schematic}
\end{figure*}

Microscope ECORE is shown in \cref{fig:schematic}(a). To mitigate the effects of vibrations on our experiment, the setup breadboard is placed on a floating, damped optical table. We also ensure that all optical elements are mounted to the table rigidly to prevent residual environmental vibrations from perturbing the beam paths. The setup is placed in an enclosure with opaque panels to reduce the amount of environmental light reaching the detector.

We use \qtyadj{510}{\nano\meter} light as PEDOT shows a strong electrochromic response at this wavelength \cite{dualcolor, prodot}. The choice of light source is also important. High-NA objectives, such as the one we use, can contain more than 10 lens elements \cite{FLM}. At the detector, small reflections from closely-spaced optical elements can interfere if the path length difference of the two interfering beams is less than the coherence length of the light. We use a superluminescent diode (SLD) (EXS210115-00, Exalos) since it has a short coherence length of \qty{\sim 8}{\micro\meter}. Therefore, parasitic reflections with a path length difference \qty{> 8}{\micro\meter} will not interfere with one another, strongly suppressing this source of technical noise. 

The light from the SLD is split into two paths using a Wollaston polarizer which can also help prevent parasitic reflections as it separates the beams by oblique angles (\qty{20}{\degree} in our case). One of these paths leads directly to one eye of a differential photodetector: this path serves as our reference beam. The remainder of the light (the signal beam) travels along the other path where it is first expanded using two plano-convex lenses to a $1/e^2$ intensity radius of \qty{\sim 2}{\milli\meter} and is then directed towards the epi-illumination port of a commercial microscope (Eclipse Ti-U, Nikon). Here, an $f=\qty{500}{\milli\meter}$ achromatic doublet lens is used to focus the light onto the back focal plane (BFP) of a 100x, 1.49-NA microscope objective (MRD01991, Nikon). The $f=\qty{500}{\milli\meter}$ lens is also used to change the angle of incidence (AOI) at the sample plane of the objective by laterally moving the focused beam at the BFP of the objective. This simplifies the alignment as changing the AOI is simply a matter of translating the lens. The objective then focuses the beam onto the sample, as shown in \cref{fig:schematic}(b). At the sample, any local changes in the electrical potential are converted into small changes in the index of refraction of the PEDOT film which result in small changes in the reflectance of the sample. Thus potential changes in the sample can be observed by monitoring the reflectance using a detector. 

The reflected beam is collected by the objective and directed towards the other eye of the differential photodetector. This homemade photodetector (as described previously in \cite{PNAS, dualcolor, prodot}) is used to increase the sensitivity of detection by reducing any common-mode noise, such as intensity noise from the light source, by manually balancing the light power impinging on both eyes of the detector. To remove high-frequency noise (and prevent aliasing in the digitized signal), \newR{the detector output is filtered using a physical low-pass filter} and then digitized with a sampling rate of \qty{10}{\kilo\hertz} and recorded \newR{(see~\cref{appendix:shotnoise})}.

\newR{Images are taken through the same objective under white-LED illumination with the signal beam blocked by a shutter; the LED is then switched off and the shutter opened for recording, separating imaging and recording by ${\ll}\qty{1}{\second}$. This keeps the LED light from degrading the ECORE signal; truly simultaneous operation could be enabled by a filter blocking all wavelengths other than \qty{510}{\nano\meter} from reaching the detector.}

We use glass coverslips ($\num{22} \times \num{22} \times \qty{0.15}{\milli\meter\tothe{3}}$) coated with indium tin oxide (ITO) (CB-90IN, Delta Technologies) as the substrates for our PEDOT devices. We attach a custom 3D printed poly(lactic acid) sample well with a circular bottom opening of diameter \qty{6}{\milli\meter} to the top of each substrate. PEDOT is then electrodeposited onto the substrate using the protocol described in \cite{dualcolor}, creating a thin film with an area of \qty{28.3}{\milli\meter\tothe{2}}. The PEDOT film thickness depends on the duration of electrodeposition -- a longer deposition time results in a thicker film.

The spatial resolution in ECORE is determined by the spot size of the beam at the sample \cite{PNAS}. \Cref{fig:schematic}(c) shows a section from an image of a typical cardiomyocyte sample recorded using a separate objective with a larger field of view. The $1/e^2$ spot sizes from various ECORE methods are also shown (superposed ellipses). When compared with the cardiomyocytes, the original ECORE beam \cite{PNAS} (large circle), and the dual-color ECORE beam \cite{dualcolor} (central ellipse) are approximately the size of whole cells or large regions of whole cells. In contrast, the microscope ECORE beam (rightmost ellipse) is smaller by a factor of 2 compared to the dual-color beam, the previously smallest ECORE beam. We are therefore able to probe the potential from smaller regions of these cardiomyocytes.

Because of the oblique incident angle, the microscope ECORE beam makes an elliptical spot on the specimen. The spot size, measured by fitting Gaussian functions to the intensity profiles of both axes of the spot, is $\num{12.9} \times \qty{9.4}{\micro\meter\tothe{2}}$ (semi-major and semi-minor axis at $\mathrm{1/e^2}$ intensity, respectively). With a typical light power of \qty{600}{\micro\watt} entering the microscope base and an objective transmittance of \qty{\sim 83}{\percent}, approximately \qty{500}{\micro\watt} of light reaches the sample. The light intensity impinging on the sample is therefore \qty{131}{\watt\per\centi\meter\tothe{2}}. If desired, the spatial resolution can be further improved to \qty{\sim 3.89}{\micro\meter} using the simple illumination scheme that we use, or to \qty{\sim 138}{\nano\meter} using annular illumination (see \cref{appendix:resolution}). In practice, however, reducing the spot size increases the light intensity at the sample, causing perturbations to the cells. This can limit the recording duration as we discuss later.

\section{Results}

\subsection{Cell-Free Characterization of Microscope ECORE}

\begin{figure}
    \includegraphics{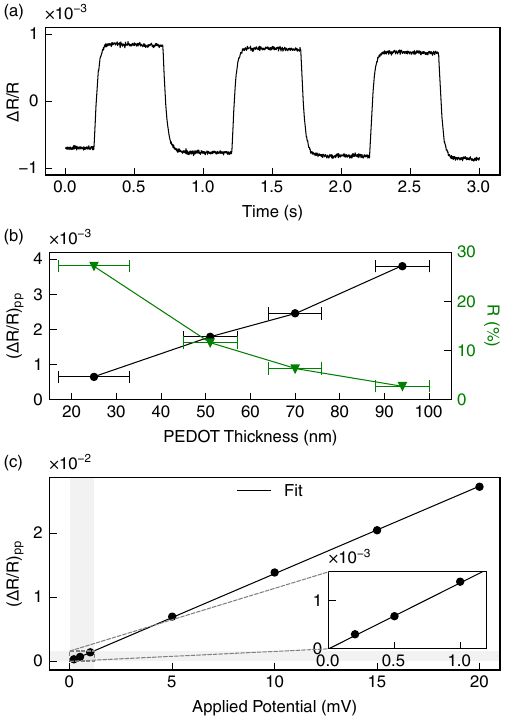}
    \caption{(a) Detector output (in units of normalized change in the reflectance of the sample $\DRR$) in response to a \qtyadj{1}{\milli\volt}, \qtyadj{1}{\hertz} applied square wave at a bias voltage of \qty{0}{\milli\volt} (with respect to a standard Ag/AgCl electrode). A \qtyadj{0.01}{\hertz} high-pass digital filter has been applied to flatten the baseline (see~\cref{appendix:shotnoise}).
    (b) \newR{The peak-to-peak amplitude $\DRRpp$ of $\DRR$} in response to a \qtyadj{1}{\milli\volt}, \qtyadj{1}{\hertz} applied square wave (at zero bias) as a function of the PEDOT film thickness (black circles and line). The reflectance $R$ of the samples is also shown (green triangles and line).
    (c) \newR{$\DRRpp$} as a function of the amplitude of the applied square wave at zero bias. The line is a linear fit to the data described by the equation \newR{$y = \num[separate-uncertainty = true]{1.3686(39)e-3}\ x/\si{mV}$}. The inset shows data points near the origin.}
    \label{fig:cellfree}
\end{figure}

To characterize our experimental apparatus, we use a potentiostat configured to apply a \qtyadj{1}{\milli\volt}, \qtyadj{1}{\hertz} square wave to a conducting ionic solution of $1 \times$ phosphate-buffered saline (PBS) in contact with the PEDOT film. We then measure the optical response by recording the output of the detector.
This output is converted into a normalized change in the reflectance of the sample, $\DRR$, where $R$ is the \newR{unperturbed} reflectance of the sample.
\newR{We use the peak-to-peak amplitude of $\DRR$ in response to the applied square wave, $\DRRpp$, as a standardized metric for evaluating a PEDOT device; a higher $\DRRpp$ value is preferable for a given applied voltage.}
A typical recording is shown in \cref{fig:cellfree}(a). This ECORE response, from a standard PEDOT film of area \qty{28.3}{\milli\meter\tothe{2}}, has a \qtyrange{10}{90}{\percent} rise time of $\tau_r = \qty[separate-uncertainty = true]{47(2)}{\milli\second}$. As previously observed \cite{PNAS, dualcolor, prodot}, $\tau_r \propto \sqrt{A_s}$, where $A_s$ is the active sensing area of a PEDOT film. This can be modeled by a resistor-capacitor circuit as noted in \cite{PNAS}. For a large cardiomyocyte with an area of \qty{2000}{\micro\meter\tothe{2}}, we expect $\tau_r = \qty{0.4}{\milli\second}$. Therefore, the films provide optical responses that are fast enough to resolve bioelectric potentials from these cells.

We studied the effect that the thickness of the PEDOT film has on \newR{$\DRRpp$}. We prepared films of varying thicknesses by performing electrodeposition for \qty{20}{\second}, \qty{40}{\second}, \qty{60}{\second}, and \qty{80}{\second}. The average film thickness of these samples was measured using a profilometer (Dektak XT-S, Bruker) to be \qty[separate-uncertainty = true]{25(8)}{\nano\meter}, \qty[separate-uncertainty = true]{51(6)}{\nano\meter}, \qty[separate-uncertainty = true]{70(6)}{\nano\meter}, and \qty[separate-uncertainty = true]{94(6)}{\nano\meter}, respectively. We find that \newR{$\DRRpp$} increases as the PEDOT film thickness is increased (black circles in \cref{fig:cellfree}(b)). However, we also find a sharper decrease in the reflectance of the samples as the film thickness increases (green triangles in \cref{fig:cellfree}(b)). Therefore, as the films gets thicker, less light reaches the detector and the SNR is reduced for the same input power. While in principle this loss of power can be compensated for by increasing the input power, in practice we cannot do so arbitrarily because high input powers can heat up the film, damaging biological cells. A good balance of high \newR{$\DRRpp$} and $R$ is found for films that are \qty{\sim 50}{\nano\meter} thick. We therefore perform all future experiments with films prepared for this thickness.

We also studied how \newR{$\DRRpp$} changes as we vary the magnitude of the applied square wave at zero bias (\cref{fig:cellfree}(c)). A linear fit to the data gives a slope of \newR{\qty[separate-uncertainty = true]{1.3686(39)e-3}{\per\milli\volt}} and accurately models the data points. The optical response of PEDOT is thus linear with the magnitude of the applied voltage, in agreement with previous experiments. Therefore, changes in the potential are accurately represented by measuring $\DRR$. Furthermore, if desired, a calibration curve such as the one shown in \cref{fig:cellfree}(c) can be used to convert the optically measured $\DRR$ into extracellular voltage levels.

\subsection{SNR and Detection Limit}

\begin{figure*}
    \includegraphics{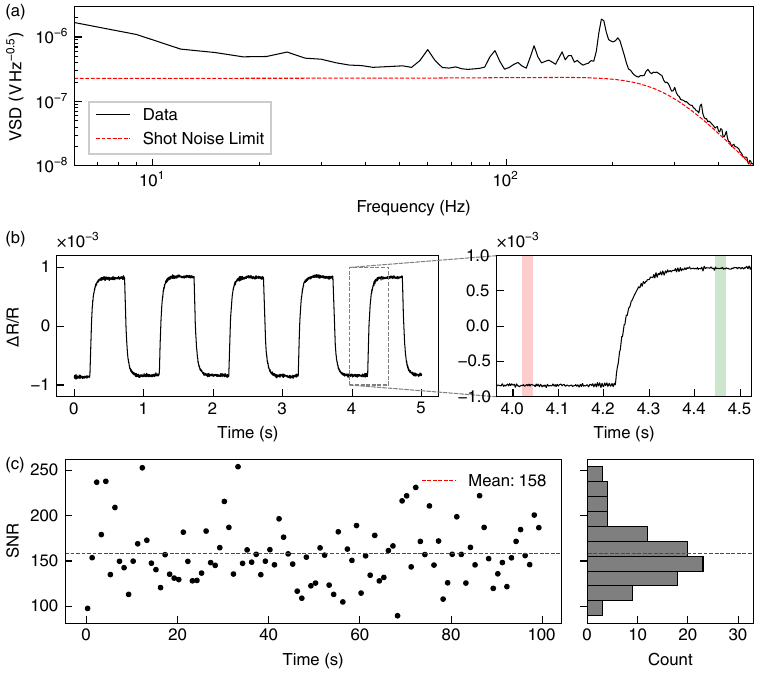}
    \caption{(a) Voltage spectral density (VSD) of the microscope ECORE signal (black trace) for \qty{600}{\micro\watt} of light entering the microscope base and \qty{44}{\micro\watt} reaching the signal eye of the detector.
    \newR{The VSD is the average of periodograms of \qtyadj{333}{ms}-long flat regions} of a \qtyadj{100}{\second}-long recording of response to a \qtyadj{1}{\milli\volt}, \qtyadj{1}{\hertz} square wave at zero bias.
    \newR{\qtyadj{0.01}{\hertz} high-pass and \qtyadj{280}{\hertz} low-pass digital filters have been applied, and the corresponding shot noise limit is shown (dashed red line, see~\cref{appendix:shotnoise}).}
    (b) The left panel shows a \qtyadj{5}{\second} section from the recording used to generate the VSD.
    A zoomed-in section is shown in the right panel and illustrates the procedure for determining the SNR of one cycle from the entire \qtyadj{100}{\second} recording.
    The red and green shaded sections show \qtyadj{25}{\milli\second} regions of the data \qty{200}{\milli\second} on the low and high levels of the square wave \newR{(part of the same regions used for the VSD)}.
    \newR{Each region is linearly detrended.
    The average of the two residual standard deviations gives the noise for this cycle, while the difference between the mean $\DRR$ of the two regions gives the peak-to-peak amplitude $\DRRpp$.}
    (c) The SNR results from all cycles in the \qtyadj{100}{\second} recording (left). The mean \newR{$\textrm{SNR} = \num{158}$} is also shown (dashed red line). A histogram of these SNR measurements is shown on the right.
    }
    \label{fig:snr}
\end{figure*}

Extracellular action potentials from cardiomyocytes have magnitudes of \qty{\sim 100}{\micro\volt} \cite{cardiac_extracellular_1, cardiac_extracellular_2}. We confirmed that the noise level of our setup is low enough to detect such small signals by characterizing a \qtyadj{100}{\second}-long microscope ECORE recording of a \qtyadj{1}{\milli\volt}, \qtyadj{1}{\hertz} applied square wave at zero bias.
\newR{The voltage spectral density (VSD) of this recording is shown in \cref{fig:snr}(a) (black trace), along with the expected shot noise limit (red dashed line).
While the VSD is within a factor of two of the shot noise limit for some of the frequency range, there are several excess noise bands, in particular around \qty{200}{Hz}.
We attribute this to vibrations of the microscope and the sample holder; in fact, we have observed some variability in frequency and amplitude of the noise bands between realignments of the setup, even when using the same sample.
Therefore, we think it likely that a more rigid setup will be able to achieve close to shot-noise-limited performance.}

To evaluate the performance of our apparatus at this noise level, we also obtained the SNR for this recording. First, we digitally apply \qtyadj{0.01}{\hertz} high-pass and \qtyadj{280}{\hertz} low-pass filters to the data \newR{(see~\cref{appendix:shotnoise}; this bandwidth leads to a 10-90\% rise time of $\qty{1.2}{ms}$, the temporal resolution that we were aiming at; it is also the same filter and cutoff as used in \cite{dualcolor} and allows for a direct comparison).} We then compute the SNR for each cycle in the \qtyadj{100}{\second} recording, as outlined in \cref{fig:snr}(b). The mean SNR for the entire recording of the \qtyadj{1}{\milli\volt} square wave is found to be \newR{\num{158}} (dashed red line in \cref{fig:snr}(c)). The detection limit is defined as the potential that gives unity SNR. We thus estimate the detection limit of microscope ECORE to be \newR{\qty{6.3}{\micro\volt}}. This result shows a successful reduction in technical noise to a level that is sufficient for detecting the typical cardiomyocyte extracellular potential. Despite the noise level being above the theoretical shot noise limit, we find that the SNR is still excellent and our detection limit is \newR{comparable to that achieved in previous demonstrations of ECORE using PEDOT films \cite{dualcolor}}.

\subsection{Recording of Cardiomyocyte Action Potentials}

\begin{figure*}
    \includegraphics{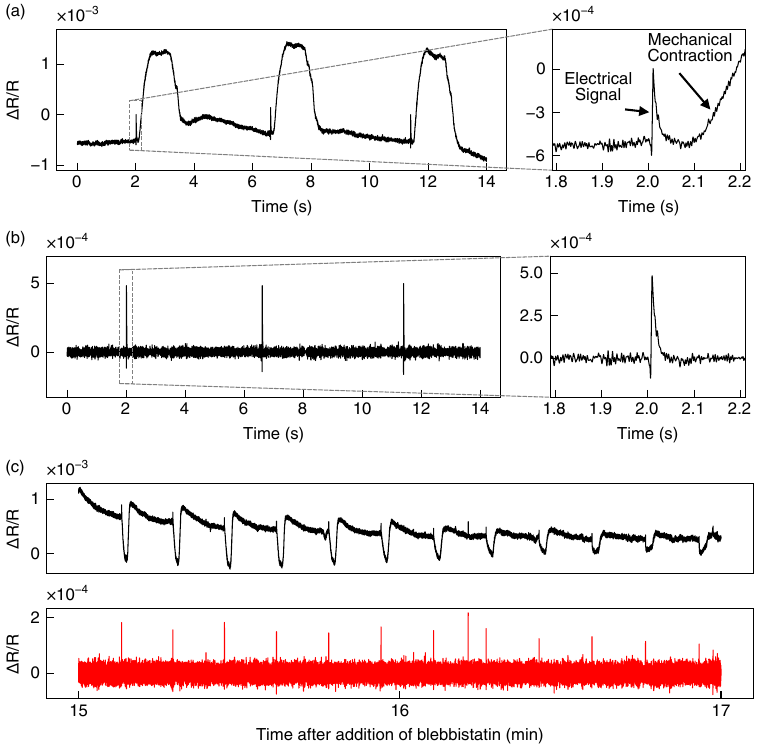}
    \caption{(a) The ECORE signal from a cardiomyocyte is shown on the left \newR{(with \qtyadj{280}{\hertz} low-pass digital filter applied, see~\cref{appendix:shotnoise})}. A zoomed-in section of the recording is shown on the right. The optical responses from the electrical signal in the cells, \newR{i.e., the extracellular action potential}, as well as the mechanical contraction are identified.
    (b) \newR{Same data after subtracting an \qtyadj{80}{ms} median baseline to remove the slow mechanical contraction signal.}
    The right panel shows the same zoom as in~(a): the electrical signal is still clearly visible but the mechanical contraction has been suppressed.
    \newR{The observed action potentials have an average SNR of \num{30} and a full-width-at-half-maximum of \qty{7.3}{ms} (the \qtyrange[range-phrase={--}]{10}{90}{\percent} rise time of \qty{1.40}{ms} is close to the lower limit set by the low-pass filter, see~\cref{appendix:shotnoise}).}
    (c) A \qty{20}{\micro\molar} solution of blebbistatin is added to a cardiomyocyte sample at $t = \qty{0}{\minute}$. The black trace shows the recorded signal, where a \qtyadj{0.1}{\hertz} high-pass first-order filter has been applied to flatten the recording baseline.
    The red trace shows the signal after removing the slow mechanical contraction signal as described in~(b).
    The black trace shows that the mechanical signal is reduced over time, while the red filtered trace confirms that the electrical signal persists.
    \newR{The low beat rate is because cell recordings were conducted at room temperature instead of \qty{37}{\celsius}, which causes stress to the cells.}
    }
    \label{fig:cardio}
\end{figure*}

We recorded the electrical activity from cultured human induced pluripotent stem cell (hiPSC)-derived cardiomyocytes using microscope ECORE (the cell culture protocol is described elsewhere \cite{prodot}).
One such recording is shown in \cref{fig:cardio}(a), \newR{where a \qtyadj{280}{\hertz} low-pass filter has been applied as done for the cell-free data (see~\cref{appendix:shotnoise}).}
As expected from previous works \cite{PNAS, dualcolor, prodot}, we find that the recording shows periodic signals associated with each beat. Zooming-in to any of the repeating features (\cref{fig:cardio}(a), right), we find a sharply peaked signal followed by a larger magnitude, longer-duration signal. We have previously confirmed that the sharp signal corresponds to the electrical signal from the cell, while the slower signal represents the mechanical contractions in the cells \cite{PNAS}. 

To focus on the electrical signals, which are of greater interest to us, we \newR{subtract an \qtyadj{80}{ms} median baseline} to suppress the signals from the mechanical contractions, as done in \cref{fig:cardio}(b).
\newR{This procedure preserves the action potential width and matches that of~\cite{dualcolor}.}
These results demonstrate that we are able to record cardiomyocyte extracellular potentials with a high SNR \newR{of up to \num{30}}.

We note that we can continuously record the action potentials from any region of a cardiomyocyte sample for at least \qty{3}{\minute}. In some experiments, beyond this point, we observe damage to the cells which manifests itself as a reduction in the electrical and mechanical signal detected through ECORE as well as a visible reduction in the cell movement observed on the camera. We attribute this relatively limited recording duration (compared to previous ECORE setups) to an increase in the light intensity at the sample owing to the smaller spot size of the beam. \newR{We confirmed that it is possible to obtain longer-term recordings by lowering the input power: We have obtained cardiomyocyte recordings for \qtyrange{6}{8}{min} at an input power of \qty{340}{\micro\watt} and at least \qty{3}{\minute} at \qty{600}{\micro\watt} (the power we usually work at, e.g., in \cref{fig:cardio}). These times decrease to \qty{0.45}{\minute} at \qty{830}{\micro\watt}.}

\newR{The performance achieved here (a detection threshold of \qty{6.3}{\micro\volt} with an input power of \qty{600}{\micro\watt}) is typical of previous ECORE implementations using PEDOT films (e.g., \qty{5.6}{\micro\volt} for the \qtyadj{561}{\nano\meter} channel of~\cite{dualcolor} with ${\sim}\qty{1}{\milli\watt}$ input power).}
\newR{A lower detection threshold has recently been demonstrated using dioxythiophene-based P(OE3)-E films~\cite{prodot}, which are in principle compatible with our setup.}
\newR{The SNR is expected to scale with the square root of the detected power $P$ if shot noise dominates, and to be independent of $P$ if the dominant noise is from mechanical vibration; across our recordings we observe $\mathrm{SNR} \propto P^{0.3}$, intermediate between the two.}
\newR{The temperature rise induced by the laser is given by the linear power density.
For example, shrinking the spot diameter twofold will require a twofold decrease in power, to keep the temperature rise the same.
We may also reduce the temperature rise by using a substrate material with improved thermal conductivity, such as sapphire.}

One significant benefit of microscope ECORE is that both the optical recording of the electrical potentials and the imaging of the specimen are done using the objective. This not only simplifies the imaging process, but also makes it easy to interact with the sample, the top of which is left completely accessible to the experimenter (\cref{fig:schematic}(b)). We took advantage of this geometry to add blebbistatin, a myosin inhibitor, to a cardiomyocyte sample in order to reduce mechanical contractions in the cells \cite{PNAS}. After the addition of blebbistatin at $t = \qty{0}{\minute}$, we monitored the ECORE signal from the sample. To prevent optical damage to the cells, we only recorded for \qty{\sim 2}{\minute} from one cardiomyocyte before recording from another cell and repeating. A cell being studied at $t \approx \qty{15}{\minute}$ shows a clear decrease in the mechanical signal strength over time (black trace in \cref{fig:cardio}(c)). When we filter out the mechanical signals altogether (red trace in \cref{fig:cardio}(c)), we find that the electrical signals persist even as the mechanical signal is reduced. This observation reconfirms that the longer, larger spikes are mechanical in nature. Furthermore, this experiment demonstrates how the access provided to the sample could be used to study the effects of drugs on biological cells.

\subsection{Recording Cardiomyocyte Action Potentials from Subcellular Regions}

\begin{figure}
    \includegraphics{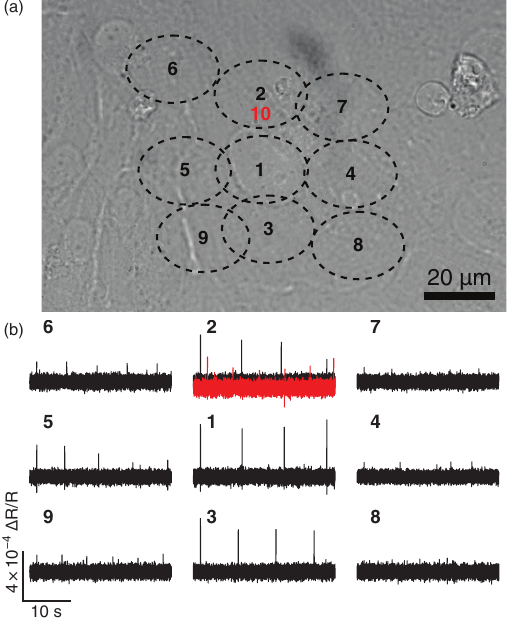}
    \caption{(a) A cardiomyocyte (center) and its neighboring cells in the field of view of the 1.49-NA objective. The dashed ellipses show the location of the probing beam. The beam was moved to a particular spot, where the ECORE signal was recorded for \qty{60}{\second} before moving to another spot and repeating this process. (b) Sections of the recordings from the corresponding ellipses in (a) after \newR{applying a \qtyadj{280}{\hertz} low-pass filter and subtracting an \qtyadj{80}{ms} median baseline (see~\cref{appendix:shotnoise})}. The recordings have been manually synchronized in time to facilitate comparison of the electrical signals in the different regions.
    \newR{The red trace (offset for visibility) is a 10th recording, a repeat of location 2 made after scanning over all positions sequentially.}
    }
    \label{fig:mapping}
\end{figure}

We next applied microscope ECORE to record the extracellular potentials from different regions of a single cardiomyocyte. In our experiments we keep the beam fixed in place and instead move the sample in any direction perpendicular to the optical axis of the objective using a translation stage. We can thus record the optical response from any user-designated region of interest by aligning the region with the known location of the beam. We did this for 9 regions of a cardiomyocyte, recording at each site for \qty{60}{\second} (dashed ellipses in \cref{fig:mapping}(a)). Sections of the resulting filtered ECORE recordings from each site are shown in \cref{fig:mapping}(b). As the recordings were performed sequentially, we have manually aligned them in time to allow comparison of the electrical signals across the different regions.
\newR{After scanning through locations 1--9, we performed a 10th recording at location 2.
Compared to the original recording 2, recording 10 shows peaks of reduced amplitude, but of larger amplitude than recordings 4, 6, 7, 8, and 9.
This shows that the spatial heterogeneity is real and not just an artifact of a changed physiological state that occurred during the recording. 
} In this cell, the electrical activity appears to be stronger along the center of the cell and weaker elsewhere. This experiment demonstrates how microscope ECORE can be used to create a map of extracellular potentials from a given cell by recording from different regions in quick succession. It also illustrates the central role played by the objective, which enables the smaller spot size and allows for the simultaneous imaging and recording of a specimen.

\newR{Overall, we tested 6 PEDOT devices with cultured cells, and used them to perform 106 recordings on 50 different cardiomyocyte cells, imaging 84 distinct regions of interest (ROIs). A majority of these recordings show action potentials, as illustrated for a particular cell in \cref{fig:mapping}(b).}

\section{Discussion}

In this work, we report electrochromic optical recordings using a high-NA objective integrated into a commercial microscope. We demonstrate recordings with signal-to-noise ratios sufficient to reach a detection limit of \newR{\qty{6.3}{\micro\volt}}. We also demonstrate recordings of action potentials from cardiomyocytes and further utilize microscope ECORE to obtain a spatial map of the action potential from different regions of one cell.

The successful reduction of noise is crucial in allowing the combination of ECORE with microscopy. This integration provides several advantages compared to previous ECORE setups. As the objective can be used for both ECORE and imaging, it is possible to observe the sample in real time and choose cells to record electrical signals from without the need for separate optics for imaging. This setup, which is also simpler to align compared to previous iterations of ECORE, leaves the top of the sample completely accessible to the experimenter. This access can be used to easily exchange drug solutions with the sample (as we have shown) or to stimulate the sample with external electrodes, for example. We also highlight the relative simplicity of the construction that is needed to build the apparatus, which requires very few components altogether and has a compact footprint. Besides the high-NA microscope objective and the microscope base, all other optical elements in the setup (\cref{fig:schematic}(a)) are inexpensive and easily obtained commercially. 

The objective also allows an improvement in the spatial resolution of ECORE. However, one barrier to further improving the spatial resolution is the reduced recording duration due to the higher light intensities associated with smaller spot sizes. This limitation may be overcome in the future with the exploration of less absorptive electrochromic materials \cite{prodot}. We also note that the recording of potentials from different sample regions is currently done sequentially (as opposed to simultaneously). In the future, we hope to combine a scanning technique (which we demonstrate in a separate manuscript under submission) with microscope ECORE to obtain simultaneous recordings from multiple points. For example, the signal beam could be scanned across different spots of the sample using an acousto-optic deflector (AOD). The AOD could be programmed to enable the recording of potentials from multiple user-designated spots simultaneously. This would allow researchers full control and spatial flexibility to study, in real time, the action potentials from a network of cells.

Microscope ECORE offers a simpler optical setup and alignment, improved spatial resolution, high recording sensitivity, and a further enhancement in the spatial flexibility over previous ECORE techniques. We therefore demonstrate microscope ECORE as a viable label-free recording method that can be utilized by researchers interested in studying bioelectric potentials, particularly those already using microscopy in their work. The introduction of ECORE into the world of microscopy not only allows for high-quality recording of extracellular action potentials, but also unlocks ECORE as an exciting tool that can be used to improve our understanding of the complex functioning of excitable cells.

\begin{acknowledgments}
We would like to thank Ravichandra Venkateshappa and Professor Joseph Wu for providing the hiPSC-cardiomyocytes. We would also like to acknowledge Professor Victoria Xu for their helpful discussions regarding microscope ECORE. Research reported in this article  was supported by the National Institute of Neurological Disorder and Stroke of the National Institutes of Health under award number 1R01NS121934 (B.C. and H.M.). This work was also supported by a Stanford Bio-X Bowes Fellowship and the NIH Stanford Graduate Training Program in Biotechnology T32GM141819 (E.L.). Y.Z. acknowledges support from the American Heart Association Career Development Award (\url{https://doi.org/10.58275/AHA.25CDA1436867.pc.gr.229603}) and The Roy J. Carver Charitable Trust Individual Investigator Award. Part of this work was performed at nano@stanford RRID:SCR\_026695.
\end{acknowledgments}

\appendix
\crefalias{section}{appendix}
\crefalias{subsection}{appendix}
\section{Spatial Resolution}
\label{appendix:resolution}

An important aspect of any microscopy technique is the spatial resolution. Here we consider the theoretical limits of the spatial resolution for microscope ECORE for two different illumination schemes.

\subsection{Spot Illumination}

We use the commonly defined resolution limit of an objective lens~\cite{axelrod_book},
\begin{align}
r = \frac{1.22\lambda_0}{2\textrm{NA}},
\end{align}
where $r$ is radius of the focused spot at the sample, $\lambda_0$ is the wavelength of the light, and $\textrm{NA}$ is the numerical aperture of the lens. With complete illumination of the aperture, the diffraction-limited spot size of a 1.49-NA objective for light with $\lambda_0 = \qty{510}{\nano\meter}$ would be $r=\qty{209}{\nano\meter}$. However, for all incident light to undergo total internal reflection (TIR) as is desired in an ECORE setup, we can only illuminate a limited part of the aperture corresponding to supercritical incident angles. 

The usable NA range is determined by the refractive indices of the four-layer structure containing the PEDOT thin film. Light enters the structure from index-matching oil ($n_0 = 1.518$) into a thin layer (\qty{<30}{\nano\meter}) of indium tin oxide ($n_1 \sim 1.84$) and a thin layer (\qty{\sim 50}{\nano\meter}) of PEDOT ($n_2 \sim 1.4$) before interacting with the last interface, i.e., water ($n_3 = 1.33$) \cite{PNAS}. For this system the critical angle is $\theta_c = \arcsin{(n_3/n_0)} = \qty{61.2}{\degree}$ and the minimum NA that must be exceeded for TIR to occur is therefore $\textrm{NA}_{\textrm{min}} = 1.33$ while the maximum NA is specified by the objective, in this case $\textrm{NA}_{\textrm{max}} = 1.49$. Typical TIR illumination is achieved by focusing light to a spot at the BFP of the objective using a lens (the $f = \qty{500}{\milli\meter}$ lens in \cref{fig:schematic}(a)). The smallest spot size at the sample that can be achieved with such illumination is therefore a focused spot at the BFP spanning the entire allowed NA range. Using the Abbe sine condition, we find that this maximal spot has $\textrm{NA} = 0.08$. Therefore, the smallest spot size that can be achieved with spot illumination is $r = \qty{3.89}{\micro\meter}$. This is considerably larger than the spot size achieved through full illumination of the aperture. If desired, we could minimize our spot size at the sample to this value by ensuring maximal illumination of the allowed range of NA values at the BFP of the objective.

\subsection{Annular Illumination}

Another approach to achieving TIR at the sample is to illuminate the entire ring (annulus) of the BFP defined by the minimum and maximum NA values for supercritical illumination. The radial intensity profile of the diffraction pattern formed at the sample plane for this illumination scheme is given by~\cite{axelrod_book}
\begin{align}
I \propto \frac{1}{{r^\prime}^2}
\left[ 
J_1\left( \frac{2\pi \textrm{NA}_{\mathrm{max}}r^\prime}{\lambda_0} \right) - 
\beta J_1\left( \frac{2\pi \textrm{NA}_{\mathrm{min}}r^\prime}{\lambda_0} \right)
\right]^2,
\end{align}
where $r^\prime$ represents the radial coordinate at the sample plane, $J_1$ is the Bessel function of the first kind of order 1, and $\beta \equiv \textrm{NA}_{\textrm{min}}/\textrm{NA}_{\textrm{max}}$. The first minimum of this profile occurs at $r^\prime = \qty{138}{\nano\meter}$, which is smaller than the first minimum location for the full aperture illumination (for which $r^\prime = \qty{209}{\nano\meter}$ as noted previously). However, as noted in \cite{axelrod_book}, more of the intensity is spread into the higher orders of the diffraction pattern. Nevertheless, annular illumination represents one way in which the spatial resolution could be further improved if desired.

\section{Filters and Shot Noise Limit}
\label{appendix:shotnoise}

\newR{\subsection{Filters}}

\newR{
The detector output passed through an analog anti-aliasing filter and a low-noise voltage amplifier (SR560, Stanford Research Systems; voltage gain of \num{100} for cardiomyocyte and \numrange{500}{10000} for cell-free data) before digitization at \qty{10}{kHz} (Digidata 1440A, Molecular Devices).
Outside the filter passband the electronic noise floor referred to the amplifier input is \qtyrange[range-phrase = { to }]{-149}{-146}{dBV/Hz}.

The anti-aliasing filter was designed as a third-order Bessel LC filter with a \qty{1}{kHz} \qty{-3}{dB} cutoff frequency in a \qty{50}{\ohm} environment.
However, an unintended impedance mismatch changed the filter's frequency response.
The mismatch caused a reduction in gain starting at ${\sim}$\qty{77}{Hz} (the \qty{46}{dB} attenuation at the Nyquist frequency nevertheless suppressed potential aliasing below the noise floor).
To correct for this, we measured the filter transfer function under the mismatched conditions using a dynamic signal analyzer, and then compensated for the filter by deconvolving in post-processing; the resulting noise amplification at high frequencies was suppressed by the subsequent digital low-pass.
The net amplification of the electronic noise floor between ${\sim}$\qty{77}{Hz} and ${\sim}$\qty{300}{Hz} was at most \qty{7.5}{dB}, and its contribution was small compared to the shot and technical noise.

The digital low-pass was a third-order Butterworth filter (\qty{280}{Hz} \qty{-3}{dB} cutoff) applied forward-backward (zero-phase, using \texttt{scipy.signal.filtfilt}~\cite{scipy}), giving a \qty{-36}{dB/octave} asymptotic roll-off and an effective \qty{-3}{dB} cutoff of \qty{242}{Hz}.
This is the same filter and the same two-pass implementation used in \cite{dualcolor,prodot}, where the forward-backward application was not stated explicitly; note that \cite{prodot} uses a cutoff of \qty{500}{Hz} (\qty{432}{Hz} effective) for their cardiomyocyte data.
The low-pass filter sets a lower limit of \qty{1.2}{ms} for the \qtyrange[range-phrase={--}]{10}{90}{\percent} rise time of the recordings.

For cell-free square wave recordings, baseline drift was removed with a first-order digital high-pass filter with a \qty{0.01}{Hz} \qty{-3}{dB} cutoff.

For cardiomyocyte recordings, the slow mechanical contraction signal was removed by subtracting an \qty{80}{ms} (800-sample) median baseline, matching the procedure of \cite{dualcolor}.
The baseline noise was taken as the standard deviation of a \qty{25}{ms} window preceding each action potential.
}

\newR{\subsection{Shot Noise Limit}}

The current spectral density of shot noise associated with a current $I$ is given by~\cite{art_of_electronics}
\begin{align}
S_I(f) = \sqrt{2eI},
\end{align}
where $e$ is the elementary charge. The current generated by a photodiode is $I = P\mathcal{R}$, where $P$ is the optical power incident on the photodiode and $\mathcal{R}$ is the responsivity of the photodiode. The photodiodes we use (FDS100, Thorlabs) have $\mathcal{R} = \qty{0.177}{\ampere\per\watt}$ for \qtyadj{510}{\nano\meter} light. For a photodetector with a feedback resistance of $R$ ($R = \qty{100}{\kilo\ohm}$ in our experiment), the voltage spectral density associated with shot noise is therefore
\begin{align}
S_V(f) = \sqrt{2eP\mathcal{R}}R.
\end{align}
This value must be multiplied by $\sqrt{2}$ to account for the fact that a differential photodetector outputs the difference of two photodiode currents.
\newR{Finally, upon passing through a filter with transfer function $G(f)$, we have
\begin{align}
S_{V,G}(f) = 2\sqrt{eP\mathcal{R}}R|G(f)|.
\end{align}
The shot noise limit shown as dashed red line in \cref{fig:snr}(a) is based on this expression, using $G(f)$ of the digital low-pass filter described in the previous section; a small contribution from the noise amplification due to the analog filter compensation has been added in quadrature in the shown line (see previous section).
In the filter passband ($|G(f)| \equiv 1$), $S_{V,G}(f) = \qty{2.24e-7}{\volt\per\hertz\tothe{0.5}}$ for the experimental parameters of \cref{fig:snr}(a) ($P = \qty{44}{\micro\watt}$).}

\newR{
Integrating over the filter yields the rms shot noise, $V_\mathrm{shot,rms} = 2\sqrt{eP\mathcal{R}}R\sqrt{B}$, where $B = \int_0^\infty |G(f)|^2\mathrm{d}f = \qty{244}{Hz}$ is the noise bandwidth of the digital low-pass.
For the parameters above, $V_\mathrm{shot,rms} = \qty{3.5}{\micro\volt}$; including the noise amplification gives an expected rms noise of $V_\mathrm{rms} = \qty{3.7}{\micro\volt}$, against the measured rms noise of \qty{9.6}{\micro\volt} for the data shown in \cref{fig:snr}(a).
}

\bibliography{bibliography}

\end{document}